\documentclass[iop]{emulateapj}
\usepackage{natbib,aas_macros}
\newcommand\Mstar{M_\star}

\newcommand\Ks{K_{\rm s}}
\newcommand\Mmin{M_{{\rm min}}}
\newcommand\sigmalogM{\sigma_{\log M}}
\newcommand\zp{z_{\rm p}}
\newcommand\Mh{\langle M_{h} \rangle}

\newcommand\Msun{M_{\odot}}
\newcommand\IC{\rm IC}
\newcommand\Ncen{N_{\rm cen}}
\newcommand\Nsat{N_{\rm sat}}
\newcommand\fsat{f_{\rm sat}}
\newcommand\Mhalo{M_{{\rm h}}}
\newcommand\Mpivot{M_{{\rm h}}^{{\rm pivot}}}
\newcommand\zspec{z_{{\rm spec}}}

\shorttitle{The Galaxy--Halo Connection in High-$z$ Universe}
\shortauthors{Ishikawa et al.}

\begin{document}

\title{The Galaxy--Halo Connection in High-Redshift Universe: Details and Evolution of Stellar-to-Halo Mass Ratios of Lyman Break Galaxies on CFHTLS Deep Fields}

\author{Shogo Ishikawa\altaffilmark{1,2,3}, Nobunari Kashikawa\altaffilmark{1,2}, Jun Toshikawa\altaffilmark{2}, Masayuki Tanaka\altaffilmark{1,2}, Takashi Hamana\altaffilmark{1,4}, Yuu Niino\altaffilmark{2}, Kohei Ichikawa\altaffilmark{2,5,6}, and Hisakazu Uchiyama\altaffilmark{1,2}}
\email{shogo.ishikawa@nao.ac.jp}

\altaffiltext{1}{Department of Astronomical Science, School of Physical Sciences, SOKENDAI (The Graduate University for Advanced Studies), Mitaka, Tokyo 181-8588, Japan}
\altaffiltext{2}{Optical and Infrared Astronomy Division, National Astronomical Observatory of Japan, Mitaka, Tokyo 181-8588, Japan}
\altaffiltext{3}{Center for Computational Astrophysics, National Astronomical Observatory of Japan, 2-21-1 Osawa, Mitaka, Tokyo 181-8588, Japan}
\altaffiltext{4}{Division of Theoretical Astronomy, National Astronomical Observatory of Japan, Mitaka, Tokyo 181-8588, Japan}
\altaffiltext{5}{Department of Physics and Astronomy, University of Texas at San
Antonio, One UTSA Circle, San Antonio, TX 78249, USA}
\altaffiltext{6}{Department of Astronomy, Columbia University,
550 West 120th Street, New York, NY 10027, USA}

\begin{abstract}
We present the results of clustering analyses of Lyman break galaxies (LBGs) at $z \sim 3$, $4$, and $5$ using the final data release of the Canada--France--Hawaii Telescope Legacy Survey (CFHTLS). 
Deep- and wide-field images of the CFHTLS Deep Survey enable us to obtain sufficiently accurate two-point angular correlation functions to apply a halo occupation distribution analysis. 
The mean halo masses, calculated as $\Mh = 10^{11.7} - 10^{12.8} h^{-1}\Msun$, increase with stellar-mass limit of LBGs. 
The threshold halo mass to have a central galaxy, $\Mmin$, follows the same increasing trend with the low-$z$ results, whereas the threshold halo mass to have a satellite galaxy, $M_{1}$, shows higher values at $z = 3 - 5$ than $z = 0.5 - 1.5$ over the entire stellar mass range. 
Satellite fractions of dropout galaxies, even at less massive haloes, are found to drop sharply from $z=2$ down to less than $0.04$ at $z=3-5$.
These results suggest that satellite galaxies form inefficiently within dark haloes at $z=3-5$ even for less massive satellites with $\Mstar < 10^{10}\Msun$. 
We compute stellar-to-halo mass ratios (SHMRs) assuming a main sequence of galaxies, which is found to provide consistent SHMRs with those derived from a spectral energy distribution fitting method. The observed SHMRs are in good agreement with the model predictions based on the abundance-matching method within $1\sigma$ confidence intervals. 
We derive observationally, for the first time, $\Mpivot$, which is the halo mass at a peak in the star-formation efficiency, at $3 < z < 5$, and it shows a little increasing trend with cosmic time at $z > 3$. 
In addition, $\Mpivot$ and its normalization are found to be almost unchanged during $0 < z < 5$. 
Our study shows an observational evidence that galaxy formation is ubiquitously most efficient near a halo mass of $\Mhalo \sim 10^{12}\Msun$ over cosmic time. 

\end{abstract}


\section{INTRODUCTION}
One of the major questions in modern astronomy is how the large-scale structure of the Universe formed and evolved. 
In the $\Lambda$CDM model, galaxies form in a clump of dark matter, termed dark haloes, the evolution of which can be well traced by cosmological $N$-body simulations \citep[e.g.,][]{springel05,prada12}. 
However, it is challenging to quantitatively link observed galaxies to their host dark haloes. 

Recent wide/deep-field multi-wavelength surveys have yielded insights into galaxy--halo connections. 
The stellar-to-halo mass ratio (SHMR), which is defined as the ratio between the dark halo mass and the galaxy stellar mass, represents the conversion efficiency from baryonic matter into stars within dark haloes. 
Wide surveys \citep[e.g.,][]{coupon12,coupon15} have been conducted to investigate the relationships between low-$z$ $(z<1)$ massive galaxies and their host dark haloes through the SHMRs, whereas deep surveys \citep[e.g.,][]{mccracken15} have provided the evolutionary history of the clustering properties and SHMRs, even for less massive galaxies $(\Mstar > 10^{9} \Msun)$ up to $z \sim 3$. 

Many theoretical and observational studies have revealed the SHMRs for galaxy populations at various redshifts. 
\citet{behroozi13a,behroozi13b} showed the evolution of SHMRs from $z = 0$ to $z = 8$ by the abundance-matching method with observational constraints and suggested that the conversion efficiency has a peak at the halo with a mass of $\sim 10^{12} \Msun$, regardless of its redshift; this result is consistent with theoretical predictions \citep[e.g.,][]{blumenthal84}. 
Recently, \citet{ishikawa15} revealed the properties of host dark haloes of star-forming galaxies at $z \sim 2$ \citep[sgzK galaxies;][]{daddi04} by precise clustering analyses by combining their own Subaru wide-field data \citep{kashikawa15} and the publicly available Canada--France--Hawaii Telescope Legacy Survey \citep[CFHTLS;][]{gwyn12} and UKIRT Infrared Deep Sky Survey \citep[UKIDSS;][]{lawrence07} data, and \citet{ishikawa16} successfully showed the decreasing trend of SHMRs for sgzK galaxies toward higher mass at $z \sim 2$. 
\citet{hildebrandt09} derived galaxy two-point angular correlation functions (ACFs) on Lyman break galaxies (LBGs) at $z = 3 - 5$ using the data of the CFHTLS Deep Fields. 
They showed a correlation between clustering strength and rest-frame UV luminosity; however, there was no significant redshift evolution of clustering strength and dark halo mass. 
\citet{barone14} firstly observed ACFs of galaxies at $z \sim 7.2$ and estimated masses of their host dark haloes using a combined galaxy samples obtained by the Hubble Space Telescope (\textit{HST}). 
\citet{harikane16} calculated the SHMRs of LBGs at $z = 4 - 7$ by combining photometric data from the \textit{HST} archives and Subaru Telescope Hyper Suprime-Cam \citep[HSC;][]{miyazaki12} early released data. 
They found that the redshift evolution of their SHMRs was consistent with the predictions via the abundance-matching technique \citep{behroozi13b}, and baryon conversion efficiency at $z \sim 4$ increased monotonically up to $\Mh \sim 10^{12} \Msun$. 

A halo occupation distribution \citep[HOD; e.g.,][]{seljak00,bullock01,berlind02,vdbosch03a,hamana04,hamana06} formalism is a useful tool to connect observed clustering signals to background dark matter distributions. 
The HOD model was developed to interpret galaxy clustering by characterizing the number of galaxies within dark haloes as probability distributions, assuming that galaxy baryonic properties are primarily governed by the masses of their host dark haloes. 
Many observational studies have successfully unveiled the relationship between observed galaxies and their host dark haloes for high-redshift Universe \citep[e.g.,][]{geach12,bethermin14,durkalec15,harikane16,ishikawa16,park16}, as well as $z < 1$ Universe \citep[e.g.,][]{zehavi11,matsuoka11,coupon12,coupon15,rodriguez16}. 

In this paper, we present the results of HOD analyses of LBGs at $z = 3$, $4$, and $5$ based upon data from the CFHTLS final data release. 
Wide and deep CFHTLS Deep Field optical photometric data enable us to collect the LBGs over a sufficiently broad range of masses to apply the HOD analysis for high-redshift Universe. 
By carrying out the precision clustering/HOD analyses for the LBGs, we can obtain clues to connect high-$z$ galaxies to their host dark haloes with various halo masses. 
We have another merit to use the data of the CFHTLS Deep Fields; one of the CFHTLS Deep Fields (D2 field) is a subset of the COSMOS field \citep{scoville07}, in which galaxy clustering patterns are reported to be different from other fields, especially at $z \sim 2$ \citep[cf.,][]{mccracken10,sato14} due to the cosmic variance \citep{clowes13,mccracken15}. 
We can eliminate the effects of the cosmic variance in angular correlation functions by summing up the results of four distinct fields. 
In addition, by combining deep near-infrared (NIR) photometric data from the WIRCam Deep Survey \citep[WIRDS;][]{bielby12}, we can derive the stellar masses, and consequently SHMRs, using a spectral energy distribution (SED) fitting technique \citep{tanaka15} based upon the multi-wavelength data. 
This enables simultaneous examination of the evolution of baryonic properties and dark haloes. 

This paper is organized as follows. 
In Section~2, we present the details of the observational data and our sample selection method. 
Stellar mass estimation of selected dropout galaxies is implemented in Section~3. 
The methods and results of the clustering analysis are presented in Section~4, discussions of these results are given in Section~5, and we conclude our analyses in Section~6. 
All magnitudes and colors are given in the AB magnitude system \citep{oke83}. 
Throughout this paper, we employ the flat lambda cosmology $(\Omega_{{\rm m}} = 0.27, \Omega_{{\rm \Lambda}} = 0.73$, and $\Omega_{{\rm b}} = 0.045)$, the Hubble constant as $h = {\rm H_{0}}/100 {\rm km} {\rm s}^{-1} {\rm Mpc}^{-1} = 0.7$, the matter fluctuation amplitude at $8h^{-1}$Mpc as $\sigma_{8} = 0.8$, and spectral index of the primordial power spectrum as $n_{{\rm s}} = 1$, which are based upon the results of the \textit{WMAP} seven-year data \citep{komatsu11} to compare with the results of previous works. 
With these assumed cosmological parameters, the ages of the Universe at $z \sim 3$, $4$, and $5$ are $\sim 1.56$, $1.12$, and $0.852$ $h^{-1}$Gyr, and physical angular scales of $1$ arcsec correspond to $5.56$, $5.03$, and $4.55$ $h^{-1}$kpc, respectively. 

\section{DATA AND SAMPLE SELECTION}
\subsection{Optical Data}
We utilized the publicly available data of the CFHTLS final data release \citep{gwyn12}. 
The CFHTLS is an optical multi-wavelength $(u^{\ast}$-, $g$-, $r$-, $i$-, and $z$-band) survey that is carried out by the CFHT MegaCam \citep{boulade03}. 
We used photometric images from the CFHTLS Deep Survey to achieve sufficient depth to collect large numbers of high-redshift galaxies to carry out accurate clustering analyses. 
The CFHTLS Deep Survey consists of four distinct fields, each of which covers $\sim 1$ deg$^{2}$. 
The mean $3\sigma$ limiting magnitudes of each optical band are $u^{\ast} = 28.07$, $g = 28.27$, $r = 27.76$, $i = 27.31$, and $z = 26.38$, respectively, with little field-to-field variation. 

\subsection{NIR Data}
Part of the survey field of the CFHTLS Deep Survey is overlapped by the WIRDS \citep{bielby12}, a wide and deep NIR survey that used the Wide-Field Infra-Red Camera \citep[WIRCam;][]{puget04} mounted at the prime focus of the CFHT. 
The WIRDS provides deep, high-quality $J$-, $H$-, and $\Ks$-band photometric images over $1.94$ deg$^{2}$, where deep optical images of the CFHTLS are available. 
We used the second data release of the WIRDS and all of the data were reduced at the CFHT and TERAPIX, as is the case for the CFHTLS. 
The mean $3\sigma$ limiting magnitudes of each NIR band are $J = 24.76$, $H = 24.65$, and $\Ks = 24.53$. 
A more detailed description of the WIRDS data can be found in \citet{bielby12}. 

\subsection{Photometry and Sample Selection}
Object detection and LBG sample selection were performed in the same manner as \citet{toshikawa16}. 
In this section, we present a brief outline of the photometry and the sample selection method. 
See also \citet{toshikawa16} for more details. 

Objects were detected by {\sc SExtractor} version 2.8.6 \citep{bertin96} in $i$-band images; other optical and NIR magnitudes were measured using ``double image'' mode. 
In addition to the total magnitude, fixed aperture ($1.4$ arcsec diameter, which was approximately twice as much as the typical seeing size of $i$-band images) photometry was also applied to measure the fluxes of objects. 
The object catalogue was limited down to $i \lesssim 27.3 $, which corresponds to a $3\sigma$ limiting magnitude in the $i$-band. 

Galaxies at $z \sim 3$, $4$, and $5$ (corresponding to $u$-, $g$-, and $r$-dropout galaxies, respectively) were selected following the color criterion developed by \citet{vdburg10} and \citet{toshikawa12} as 
\begin{eqnarray}
u{\rm -dropouts}: &\Bigl(&1.0<(u^{\ast}-g)\Bigr) \cap \Bigl(-1.0<(g-r)<1.2\Bigr) \nonumber \\
&\cap& \Bigl(1.5\times(g-r)<(u^{\ast}-g)-0.75\Bigr), 
\label{eq:udrop}
\end{eqnarray}
\begin{eqnarray}
g{\rm -dropouts}: &\Bigl(&1.0<(g-r)\Bigr) \cap \Bigl(-1.0<(r-i)<1.0\Bigr) \\
&\cap& \Bigl(1.5\times(r-i)<(g-r)-0.80\Bigr) \cap \bigl(u^{\ast}>m_{u^{\ast}, 2\sigma}\bigr), \nonumber
\label{eq:gdrop}
\end{eqnarray}
and
\begin{eqnarray}
r{\rm -dropouts}: &\Bigl(&1.2<(r-i)\Bigr) \cap \Bigl(-1.0<(i-z)<0.7\Bigr) \\
&\cap& \Bigl(1.5\times(i-z)<(r-i)-1.00\Bigr) \cap \bigl(g>m_{g, 2\sigma}\bigr). \nonumber
\label{eq:rdrop}
\end{eqnarray}
It is noted that $m_{u^{\ast}, 2\sigma}$ and $m_{g, 2\sigma}$ represent the $2\sigma$ limiting magnitudes of the $u^{\ast}$- and $g$-bands. 
The total number of $u$-, $g$-, and $r$-dropout galaxies selected by the above criterion for the entire CFHTLS Deep Fields were $63,563$, $47,760$, and $9,477$, respectively. 
The effective survey area of the CFHTLS Deep Fields was $3.38$ deg$^{2}$ after masking the regions around saturated objects and frame edges. 
Refer to \citet{toshikawa16} for the validity and the sample completeness of the selection methods. 

\section{STELLAR MASS ESTIMATION}
In this section, we estimate the stellar mass of the dropout galaxy samples in two ways, using $1)$ an SED fitting technique and $2)$ a main sequence of star-forming galaxies. 

\subsection{SED Fitting}
By combining the photometric images of the CFHTLS and the WIRDS, five optical data and three NIR data are available and we can apply an SED fitting technique to derive the photometric redshift as well as the stellar mass of each LBG sample. 
We use an SED fitting code with Bayesian physical priors, {\sc Mizuki} \citep{tanaka15}. 
The SED fitting is only applied for galaxies detected in NIR images to ensure independence of their stellar-mass estimation from those by the main sequence of star-forming galaxies. 

Galaxy SED templates are generated by the spectral synthesis model of \citet{bc03}. 
An exponential-decay model with varying declination time-scale, $\tau$, is assumed for the star-formation history of the galaxy template. 
The SED templates are only considered for solar metallicity abundance. 
We confirm that the stellar masses and photometric redshifts are not significantly changed when including the SED templates with sub-solar abundance. 
It is also assumed that the initial mass function (IMF) is a Chabrier IMF \citep{chabrier03}, the dust attenuation follows the Calzetti curve with varying the optical depth, $\tau_{V}$ \citep{calzetti00}, and the IGM attenuation follows the relation of \citet{madau95}. 
Nebular emission lines are added to the SED templates of \citet{bc03} with the intensity ratios of \citet{inoue11} and the differential dust extinction law proposed by \citet{calzetti97}. 

Photometric redshifts are evaluated through the likelihood: 
\begin{equation}
\mathcal{L} \propto \exp{\left(-\chi^{2}_{{\rm SED fit}}/2 \right)}, 
\label{eq:likelihood}
\end{equation}
where the $\chi^{2}_{{\rm SED fit}}$ can be computed as 
\begin{equation}
\chi^{2}_{{\rm SED fit}} = \sum_{i} \frac{\left(f_{i, {\rm obs}} - \alpha f_{i, {\rm model}}\right)^{2}} {\sigma_{i, {\rm obs}}^{2}}. 
\label{eq:SEDfit}
\end{equation}
$f_{i, {\rm obs}}$ and $f_{i, {\rm model}}$ are the observed and the model SED fluxes of the $i$-th filter, and $\sigma_{i, {\rm obs}}$ is the uncertainty of the $i$-th observed flux. 
$\alpha$ is a normalization parameter that controls the amplitude of the model SED. 
Physical priors are multiplied by the likelihood to obtain posteriors. 
It should be noted that the results of the SED fitting (e.g., the redshift distribution and the stellar-mass distribution) do not significantly change by putting off the physical priors. 
We evaluate the photometric redshifts of $17,341$ $u$-dropout galaxies and $13,298$ $g$-dropout galaxies, respectively. 

\begin{figure}[tbp]
\begin{center}
\epsscale{1.2}
\plotone{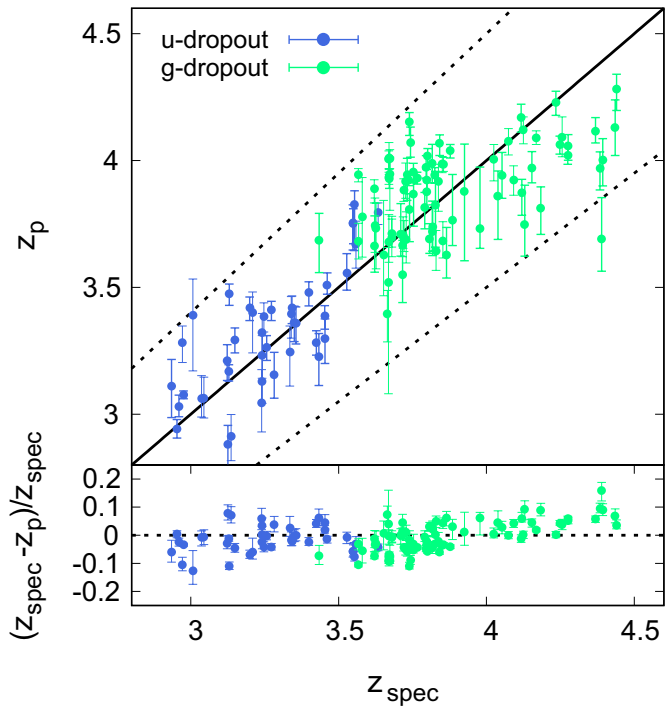}
\caption{Top panel shows a comparison between the photometric redshift and the spectroscopic redshift. Blue and green points indicate the $u$- and $g$-dropout galaxies. The solid black line is a one-to-one correspondence and the dotted black lines represent $|z_{{\rm spec}} - \zp|/(1+\zspec) = \pm 0.10$. Bottom panel shows the relative error between the photometric redshift and the spectroscopic redshift as a function of the spectroscopic redshift. }
\label{fig:comp_redshift}
\end{center}
\end{figure}

A part of our dropout sample have been implemented spectroscopic observations by \citet{toshikawa16}; we compare the photometric redshift, $\zp$, with the spectroscopic redshift, $\zspec$, to check its accuracy. 
Figure~\ref{fig:comp_redshift} is a $\zspec$ verses $\zp$ diagram of $u$- and $g$-dropout galaxies. 
The numbers of spectroscopic samples are $42$ ($u$-dropout galaxies) and $83$ ($g$-dropout galaxies), respectively. 
The photometric redshifts show good agreement with the spectroscopic redshifts; it is improved a reliability of the results of our SED fitting. 

\begin{figure*}[tbp]
\begin{center}
\epsscale{1.13}
\plotone{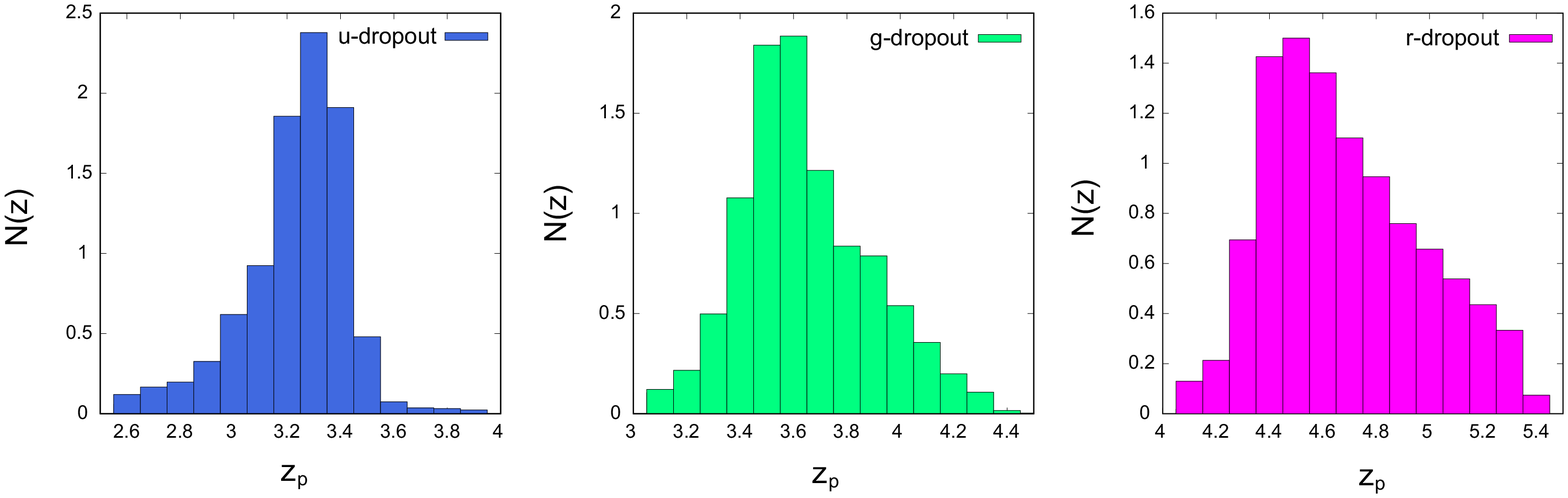}
\caption{Normalized photometric-redshift distributions of $u$- (left), $g$- (center), and $r$-dropout galaxies (right) with stellar masses of $\log(\Mstar/\Msun)>10.0$. Photometric redshifts and stellar masses are evaluated using the spectral energy distribution (SED) fitting code with physical priors, {\sc Mizuki} \citep{tanaka15}. The means and standard deviations of these redshift distributions are $z_{p} = 3.11 \pm 0.32$, $z_{p} = 3.62 \pm 0.28$, and $z_{p} = 4.67 \pm 0.32$, respectively. }
\label{fig:Nz}
\end{center}
\end{figure*}

Figure~\ref{fig:Nz} is the photometric-redshift distributions of $u$-, $g$-, and $r$-dropout galaxies. 
The means and standard deviations of these redshift distributions are $z_{p} = 3.11 \pm 0.32$, $z_{p} = 3.62 \pm 0.28$, and $z_{p} = 4.67 \pm 0.32$, respectively. 
These distributions are approximately consistent with the results of \citet{toshikawa16}, who calculated the redshift distributions of $u$-, $g$-, $r$-, and $i$-dropout galaxies using a mock LBG catalogue generated by \citet{bc03} SED models, and the photometric redshifts of \citet{hildebrandt09} using only the optical images of the CFHTLS. 

\subsection{Main Sequence of Star-Forming Galaxies}
The SED fitting method allows estimation of the plausible stellar mass; however, the total sample number decreases due to the limited survey area in which NIR data are available. 
Therefore, we also utilize a ``main sequence'' of star-forming galaxies \citep[MS; e.g.,][]{daddi07,rodighiero11,koyama14} to evaluate galaxy stellar masses, instead of the SED fitting results, to use all of the LBG samples obtained in the entire CFHTLS field and achieve high S/N clustering analyses. 
The MS relation is a tight correlation between galaxy stellar masses $(\Mstar)$ and star-formation rates (SFRs) for star-forming galaxies. 
The SFRs of our LBGs are converted from their rest-frame $1450{\rm \AA}$ luminosities using the relation proposed by \citet{kennicutt98}. 
The power-law slope of the rest-frame UV continuum, $\beta$, is measured by $(g-r)$, $(r-i)$, and $(i-z)$ colors for $u$-, $g$-, and $r$-dropout galaxies, and $1450{\rm \AA}$ luminosities are determined by extrapolating from $r$-, $i$-, or $z$-band magnitudes assuming $\beta$. 
Dust extinction of UV flux is corrected by assuming the dust extinction low by \citet{calzetti00}. 

We adopt a simple linear correlation for the MS as 
\begin{equation}
{\rm SFR}\left(z, M_{\star}\right) = A\left(z\right) \times \frac{M_{\star}}{10^{11} \Msun} \Msun{\rm yr^{-1}},
\label{eq:MS}
\end{equation}
which is the same relation that \citet{tanaka15} put as a prior for their SED fitting technique. 
$A(z)$ is a redshift-dependence term defined as 
\begin{eqnarray}
A(z) = \left\{
\begin{array}{c}
10 \times (1+z)^{2.1} \ \ ( z  <   2 ) \\
19 \times (1+z)^{1.5} \ \ (z  \geq  2 ).
\end{array}
\right .
\label{eq:MS_z}
\end{eqnarray}
The photometric redshift at the peak of its distribution is adopted when we compute the stellar masses using the MS relation. 
We introduce $\pm 0.2$ dex scatter in the relation between the SFR and the stellar mass \citep[e.g.,][]{speagle14}. 
Observational results \citep[e.g.,][]{magdis10,salmon15,alvarez16} and smoothed particle hydrodynamics simulations \citep[e.g.,][]{katsianis15} support that dropout galaxies at $z = 3$, $4$, and $5$ follow our assumed MS relation. 
We compare stellar-mass functions of each dropout galaxy with the results of \citet{santini12} and \citet{song16}, and the lowest stellar-mass limit is determined as the mass of which the observed stellar-mass functions reach $\sim 70\%$ completeness. 
The number of LBG sample of each subsample are summarized in Table~\ref{tab:param}. 

\subsection{Consistency of the Stellar Mass Estimation Between the SED Fitting and the MS Relation}
We compute stellar masses of dropout galaxy samples with two independent estimates: the main sequence of star-forming galaxies and the SED fitting technique. 
The MS relation is a convenient way to assess stellar masses of star-forming galaxies from their UV luminosities; however, derived stellar masses could suffer from non-negligible uncertainties due to the relatively large scatter. 
On the other hand, the SED fitting technique is frequently used to give more reliable estimates of the stellar mass, although broad wavelength coverage of the data set is required. 

The Balmer/$4000{\rm \AA}$ break is an essential spectral feature to obtain accurate stellar masses in the SED fitting technique. 
It should be noted that the Balmer break can be traced by WIRCam data for $u$- and $g$-dropout galaxies, but not for $r$-dropout galaxies. 

\begin{figure}[tbp]
\begin{center}
\epsscale{1.2}
\plotone{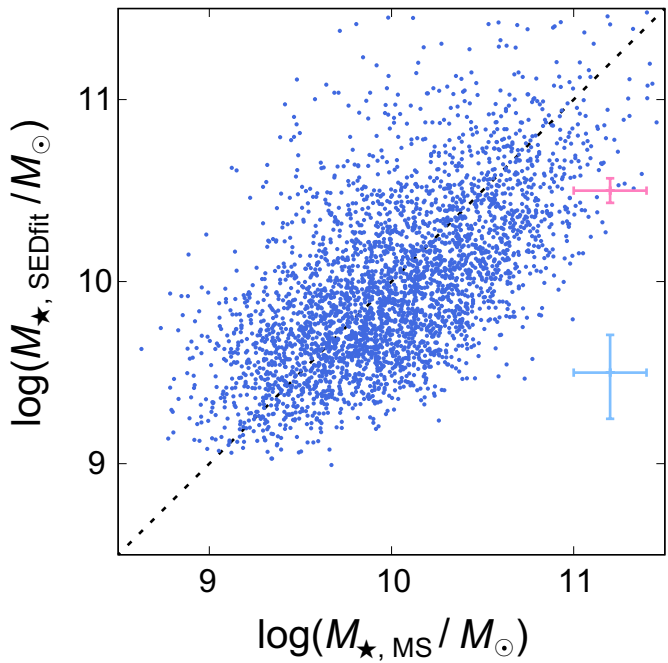}
\caption{Comparison of stellar masses of $u$-dropout galaxies in the D1 field derived by the SED fitting technique and assuming the main sequence of star-forming galaxies. The black dotted line represents the relation of the one-to-one correspondence. Crosses shown in the figure at the right represent the typical errors of $\log(M_{\star, {\rm SEDfit}}/\Msun) < 10$ (blue) and $\log(M_{\star, {\rm SEDfit}}/\Msun) \geq 10$ (red) samples. }
\label{fig:comp_Mstar}
\end{center}
\end{figure}

Figure~\ref{fig:comp_Mstar} is a comparison of stellar masses estimated using the SED fitting technique and the MS relation for $u$-dropout galaxies in the D1 field ($3,623$ galaxies). 
These two estimations show nearly a one-to-one correspondence, albeit with relatively large scatter. 
Scatters on the stellar mass evaluated from the MS relation is due to the intrinsic $\pm 0.2$ dex scatter in equation~(\ref{eq:MS}), whereas the small scatter of the SED fitting technique for the massive galaxies (red cross at the right of Figure~\ref{fig:comp_Mstar}) originates from their apparent Balmer/$4000{\rm \AA}$ break. 
The same consistency can also be obtained for $g$-dropout galaxies. 
We assume that these two estimations are consistent with respect to the other, with minimal significant difference. 
Hereafter, we will use the MS relation that allows the stellar-mass estimation down to the faint magnitudes for the entire CFHTLS field, even without WIRCam data, to estimate the stellar mass in the following analyses. 
We also assume that the consistency between these two estimations is valid for $r$-dropout galaxies. 
The effects of these two stellar-mass estimation on the SHMR results are discussed in Section~5.4.1. 

\section{CLUSTERING AND HOD ANALYSIS}
\subsection{Method}
\subsubsection{the galaxy two-point angular correlation function}
ACF, $\omega(\theta)$, is a good physical quantity for statistically estimating the galaxy clustering strength \citep[e.g.,][]{totsuji96,peebles80}. 
We employ the estimator proposed by \citet{landy93} as 
\begin{equation}
\omega \left(\theta\right) = \frac{{\rm DD} - {\rm 2DR} + {\rm RR}}{{\rm RR}}, 
\label{eq:acf}
\end{equation}
where DD, DR, and RR are the normalized pair counts of galaxy--galaxy, galaxy--random, and random--random samples, respectively, with a separation angle of $\theta \pm \delta\theta$ in degree units. 
It is known that the ACF of \citet{landy93} is biased due to the limitation of the survey field, termed the integral constraint, $\IC$. 
The integral constraint can be estimated as 
\begin{equation}
\IC = \frac{\Sigma \theta^{1-\gamma} {\rm RR}(\theta)}{\Sigma {\rm RR}(\theta)},
\label{eq:ic}
\end{equation}
assuming that the ACF is obtained by a power-law form as $\omega(\theta) = A_{\omega}\theta^{1-\gamma}$ \citep{roche99}, where $A_{\omega}$ is the amplitude of the ACF. 
We correct the effect of the integral constraint as follows:
\begin{equation}
\omega(\theta) = \omega_{{\rm meas}}(\theta) \times \frac{\theta^{1-\gamma}}{\theta^{1-\gamma} - \IC},
\label{eq:cor_acf}
\end{equation}
where $\omega_{{\rm meas}}(\theta)$ is a measured ACF. 

We also correct the effect of contaminations from low-$z$ galaxies, which decrease the measured clustering amplitudes. 
We assume that contamination sources have homogeneous distributions, and contamination effects are corrected by multiplying clustering amplitudes by $1/(1-f_{{\rm c}})^{2}$, where $f_{{\rm c}}$ is the contamination fraction calculated by the photometric-redshift distributions. 
It should be noted that the effect of contamination correction is much smaller than the statistical jackknife error of each angular bin. 

\subsubsection{the HOD modeling}
The HOD model assumes that the number of galaxies within a dark halo depends only on the masses of the dark haloes. 
We assume a standard halo occupation function proposed by \citet{zheng05} as 
\begin{equation}
N_{{\rm tot}}(M) = \Ncen(M) \times \left[ 1 + \Nsat(M)\right],
\label{eq:hof}
\end{equation}
where $N_{{\rm tot}}(M)$, $\Ncen(M)$, and $\Nsat(M)$ are the numbers of total, central, and satellite galaxies within a dark halo with a mass of $M$, respectively. 
The central galaxy occupation, $\Ncen(M)$, is expressed as 
\begin{equation}
\Ncen(M) = \frac{1}{2} \left[ 1 + {\rm erf} \left( \frac{\log M - \log \Mmin}{\sigmalogM}\right) \right],
\label{eq:central}
\end{equation}
and the satellite galaxy occupation, $\Nsat(M)$, is given by 
\begin{equation}
\Nsat(M) = \left(\frac{M - M_{0}}{M_{1}} \right)^{\alpha}. 
\label{eq:satellite}
\end{equation}
In this HOD model, galaxy occupation is tuned by five free parameters: $\Mmin$, $M_{1}$, $M_{0}$, $\alpha$, and $\sigmalogM$. 

In predicting ACFs from the HOD model, we assume the halo mass function (HMF) proposed by \citet{sheth99}, the halo bias factor of \citet{tinker10}, and the density profile of the dark halo as an NFW profile \citep{navarro97}. 
Halo mass--concentration relation is employed the model of \citet{takada03}. 
Photometric redshifts of LBG samples are employed as their redshift distributions (see Section~3.1.). 
We use the analytical formula of HMF of \citet{sheth99} since it is easy to treat the fitting function and to compare our results with literature; however, recent cosmological $N$-body simulations suggest that the HMF of \citet{sheth99} overpredicts the massive end at high redshift $(z \gtrsim 2)$. 
\citet{ishiyama15} calculated the redshift evolution of the HMF using the result of the $\nu^{2}$GC (New Numerical Galaxy Catalog) simulation \citep{makiya16} and found that their best-fitted HMF at high-mass end $(\Mhalo \gtrsim 10^{13}\Msun)$ at $z>3$ is lower than \citet{sheth99}. 
However, the difference is small within their $1\sigma$ errors even at least $z=7$. 
\citet{tinker08} and its improved model tuned to high-$z$ galaxies by \citet{behroozi13b} showed that, at high redshift, massive end of the HMF of \citet{sheth99} overpredicts compared to the results of the $N$-body simulations. 
However, the difference of the HMF at $z=3$ between \citet{sheth99} and \citet{behroozi13b} is negligible for less-massive haloes ($\Mhalo \sim 10^{12}\Msun$), whereas the apparent excess ($\sim 70\%$ excess of the \citeauthor{sheth99} HMF) can be seen for massive haloes $(\Mhalo \sim 10^{14}\Msun)$, which is a typical halo mass range of $M_{1}$ (see Table~\ref{tab:param}). 
However, both HMFs are within the amplitudes difference at $1\sigma$ error range of $M_{1}$; thus, we conclude that the difference of the HMF does not make a large impact on the final results. 

\subsection{Results}
\begin{figure}[tbp]
\begin{center}
\epsscale{1.0}
\plotone{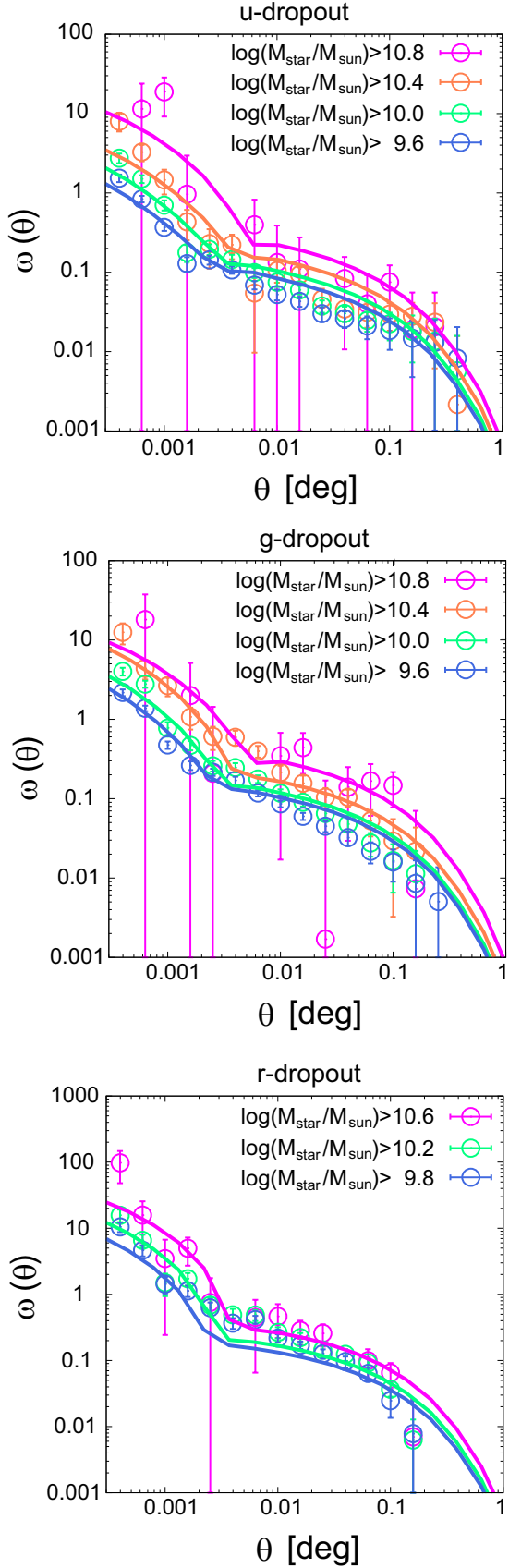}
\caption{Observed angular correlation functions (ACFs) of $u$- (top), $g$- (middle), and $r$-dropout galaxies (bottom) as a function of separation angular scales in degree units. The dropout galaxy samples are divided by their stellar masses estimated by assuming the main sequence of star-forming galaxies. Error bars of each ACF are evaluated by the jackknife resampling technique. Solid lines are the best-fit ACFs of each subsample calculated by the halo occupation distribution (HOD) model. Correlations between angular bins are taken into consideration via covariance matrices computed by the jackknife resampling method. }
\label{fig:HOD}
\end{center}
\end{figure}

To investigate the dependence of clustering properties of LBGs on their stellar masses, we divide our samples into subsamples by their stellar masses. 
We present the ACFs of each dropout galaxy in Figure~\ref{fig:HOD}. 
Error bars of each angular bin are evaluated by a jackknife resampling method \citep[cf.,][]{norberg09}. 
The ACFs show that the dropout galaxies are more strongly clustered with higher stellar mass thresholds. 
This trend is also seen in previous clustering studies with stellar-mass limit samples at lower-$z$ \citep[e.g.,][]{bethermin14,mccracken15}. 

We adopt the HOD framework assuming the halo model as shown in Section~4.1.2. 
We use a publicly available code, ``CosmoPMC'', to constrain the HOD parameters and evaluate the best-fit ACFs \citep{wraith09,kilbinger10}. 
The HOD parameters are estimated by a Markov Chain Monte Carlo (MCMC) simulation. 
For each step, we calculate the ACF from the HOD model, $\omega_{{\rm HOD}} (\theta)$, and compute $\chi^{2}$ by comparing the observed ACF, $\omega_{{\rm obs}} (\theta)$, and abundance, as follows: 
\begin{eqnarray}
\chi^{2} & = & \sum_{i, j} \Bigl( \omega_{{\rm obs}} (\theta_{i}) - \omega_{{\rm HOD}} (\theta_{i}) \Bigr) (C^{-1})_{ij} \Bigl( \omega_{{\rm obs}} (\theta_{j}) - \omega_{{\rm HOD}} (\theta_{j}) \Bigr) \nonumber \\
& + & \frac{[n^{{\rm obs}}_{g} - n^{{\rm HOD}}_{g}]^{2}}{\sigma^{2}_{n_{g}}},
\label{eq:chi}
\end{eqnarray}
where $(C^{-1})_{ij}$ is an $(i, j)$ element of an inverse covariance matrix, $n^{{\rm obs}}_{g}$ and $n^{{\rm HOD}}_{g}$ are the number densities of galaxies from observation and the HOD model, and $\sigma_{n_{g}}^{2}$ is the statistical $1\sigma$ error of $n^{{\rm obs}}_{g}$. 
We note that both a Poisson error and the cosmic variance are considered in assessing $\sigma_{n_{g}}^{2}$ \citep{trenti08}. 
$n^{{\rm obs}}_{g}$ is calculated by
\begin{equation}
n^{{\rm obs}}_{g} = \frac{N_{{\rm gals}}}{\Omega \int_{z} dz q(z) \frac{dV}{dz}}, 
\label{eq:ngobs}
\end{equation}
where $N_{{\rm gals}}$ is the total number of galaxies, $\Omega$ is the solid angle of the observation field, $q(z)$ is normalized redshift distribution at $z$, and the and $\frac{dV}{dz}$ is a comoving volume element. 
$n^{{\rm HOD}}_{g}$ can be calculated using the best-fit halo occupation function as 
\begin{equation}
n^{{\rm HOD}}_{g} = \int dM n(M) N_{tot}(M). 
\label{eq:nghod}
\end{equation}

The covariance matrix is estimated by the ``delete-one'' jackknife method \citep{shao89}. 
We divide our survey field into $64$ sub-fields and evaluate the ACF by excluding one sub-field. 
This procedure is repeated $64$ times. 
The covariance matrix is calculated as follows: 
\begin{equation}
C_{ij} = \frac{N-1}{N} \sum_{k=1}^{N} \left( \omega_{k}(\theta_{i}) - \overline{\omega}(\theta_{i}) \right) \times \left( \omega_{k}(\theta_{j}) - \overline{\omega}(\theta_{j}) \right),
\label{eq:covariance}
\end{equation}
where $\overline{\omega}(\theta_{i})$ is the averaged ACF of the $i$-th angular bin. 
We introduce the correlation factor of \citet{hartlap07} to obtain an unbiased inverse covariance matrix. 

The results of the HOD analyses are shown over the observed ACFs in Figure~\ref{fig:HOD}. 
We note that $\sigmalogM$ and $\alpha$ are fixed as $\sigmalogM = 0.30$ and $\alpha = 1.0$ to achieve better constraints of HOD mass parameters, especially for $\Mmin$ and $M_{1}$. 
The excesses for ACFs from a single power law at small-angular scales are well described by the HOD model. 
The best-fit HOD parameters and deduced parameters are presented in Table~\ref{tab:param}. 

\begin{figure}[tbp]
\begin{center}
\epsscale{1.20}
\plotone{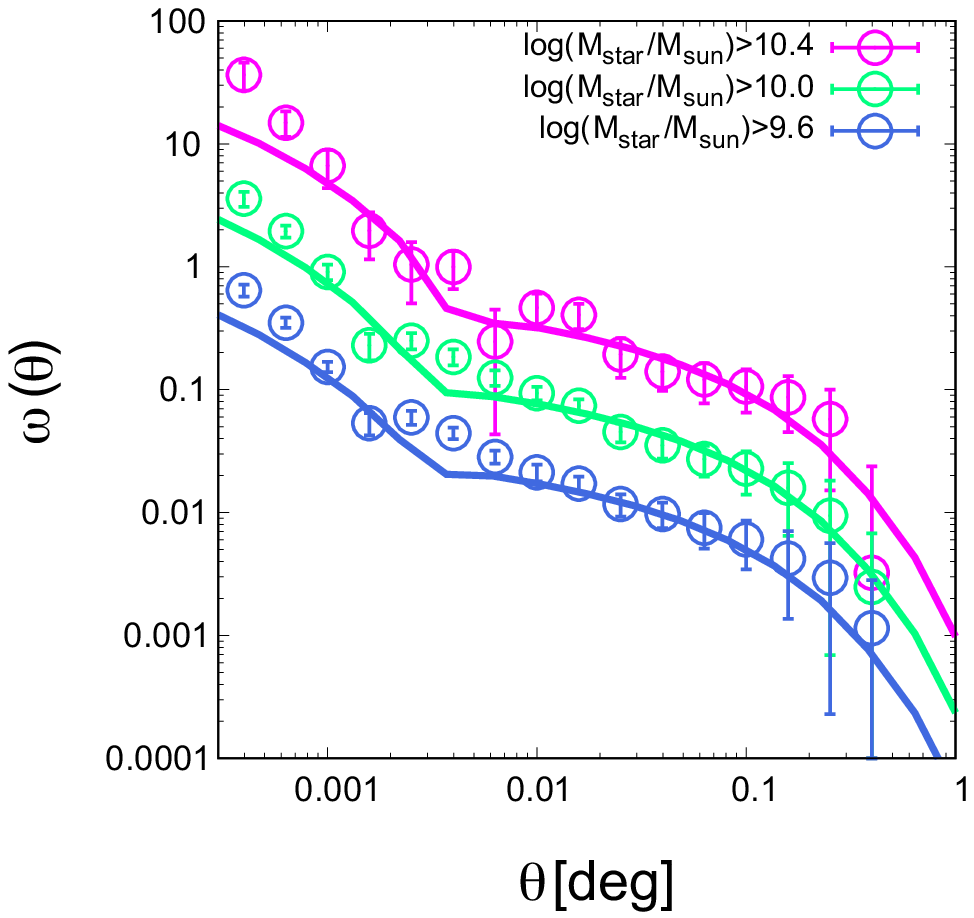}
\caption{Results of the HOD fitting of $u$-dropout galaxies by varying all of the HOD free parameters, $\Mmin$, $M_{1}$, $M_{0}$, $\sigmalogM$, and $\alpha$. To clarify the results, amplitudes of the observed ACFs and the HOD fittings are scaled arbitrarily. }
\label{fig:HOD_5free}
\end{center}
\end{figure}

Our HOD fittings slightly deviate from the observed ACFs at the $2$-halo term, i.e., $0.01 \lesssim \theta \lesssim 0.1$ for $u$- and $g$-dropout galaxies. 
This could be due to the fixing HOD parameters, especially $\sigmalogM$ that controls the occupation of the central galaxy near the dark halo mass of $\Mmin$. 
We check whether the HOD fitting will be improved by varying all of the HOD free parameters for $u$-dropout samples and the results are shown in Figure~\ref{fig:HOD_5free} and Table~\ref{tab:param_fix}. 
We note that the two most massive subsamples, i.e., $\log{(\Mstar/\Msun)} \geq 10.8$ and $\log{(\Mstar/\Msun)} \geq 11.0$ bins, have not enough S/N ratios to fit the HOD model with five free parameters. 
By varying $\sigmalogM$ and $\alpha$, fittings of the $2$-halo terms seem to be improved; however, the ACF of the $1$-halo term at small scales and the transition scales between the $1$-halo term and the $2$-halo term cannot be reproduced. 
In addition, values of the $\chi^{2}$ are not significantly changed by varying $\sigmalogM$ and $\alpha$ due to the decrease of the degree-of-freedom. 
Moreover, all of the fitting results show very small $\sigmalogM$, i.e., $\sigmalogM<0.1$, indicating that the occupation of central galaxies follows almost the step function. 
This could imply that the halo occupation function at $z>3$ is not the same as that of local Universe. 
The validity of the halo occupation function in the form of equation~(\ref{eq:central}) and (\ref{eq:satellite}) should be verified for high-$z$ galaxies; however, the discussion is beyond the scope of this paper. 
In this study, we adopt the results of the HOD fitting by fixing the parameters of $\sigmalogM=0.30$ and $\alpha=1.0$ to confine the following discussion to $\Mmin$ and $M_{1}$. 

\begin{turnpage}
\begin{table*}[htb]
\begin{center}
\caption{The Number of Dropout Samples and the Best-Fit HOD Parameters with $1\sigma$ errors of Each Subsample Limited by the Stellar Mass with Fixing $\sigmalogM = 0.30$ and $\alpha = 1.0$}
\begin{tabular}{lcccccccc} \hline \hline
 & $\log(M_{{\rm \star, limit}}/\Msun)$ & $N$ & $\log(\Mmin/h^{-1}\Msun)$ & $\log(M_{1}/h^{-1}\Msun)$ & $\log(M_{0}/h^{-1}\Msun)$ & $\log(\Mh/h^{-1}\Msun)$ & $\fsat$ & $\chi^{2}/{\rm dof}$ \\ \hline
$u$-dropout & $9.4$ & $59,233$ & $11.39^{+0.02}_{-0.02}$ & $13.17^{+0.06}_{-0.06}$ & $7.98^{+1.36}_{-1.37}$ &  $11.79^{+0.02}_{-0.02}$ & $0.037 \pm 0.003$ & $6.08$ \\ 
 & $9.6$ & $50,159$ & $11.49^{+0.03}_{-0.03}$ & $13.34^{+0.05}_{-0.05}$ & $8.01^{+1.35}_{-1.35}$ & $11.86^{+0.01}_{-0.01}$ & $0.031 \pm 0.002$ & $6.90$ \\ 
 & $9.8$ & $36,890$ & $11.60^{+0.03}_{-0.03}$ & $13.48^{+0.06}_{-0.05}$ & $7.97^{+1.34}_{-1.35}$ & $11.94^{+0.01}_{-0.01}$ & $0.027 \pm 0.002$ & $5.79$ \\
 & $10.0$ & $23,666$ & $11.69^{+0.03}_{-0.03}$ & $13.54^{+0.06}_{-0.06}$ & $7.98^{+1.35}_{-1.36}$ &  $12.00^{+0.01}_{-0.02}$ & $0.027 \pm 0.002$ & $5.85$ \\
 & $10.2$ & $13,081$ & $11.83^{+0.03}_{-0.02}$ & $13.68^{+0.07}_{-0.06}$ & $7.98^{+1.37}_{-1.40}$ &  $12.11^{+0.01}_{-0.01}$ & $0.026 \pm 0.002$ & $5.22$ \\
 & $10.4$ & $5,901$ & $11.93^{+0.02}_{-0.02}$ & $14.04^{+0.11}_{-0.10}$ & $8.53^{+2.42}_{-2.44}$  &  $12.18^{+0.02}_{-0.02}$ & $0.014 \pm 0.002$ & $5.42$ \\
 & $10.6$ & $2,230$ & $12.12^{+0.03}_{-0.03}$ & $14.28^{+0.15}_{-0.10}$ & $8.49^{+2.46}_{-2.35}$ & $12.33^{+0.01}_{-0.01}$ & $0.011 \pm 0.002$ & $2.63$ \\
 & $10.8$ & $664$ & $12.42^{+0.04}_{-0.03}$ & $14.43^{+0.26}_{-0.25}$ &  $8.56^{+2.37}_{-2.38}$ & $12.55^{+0.02}_{-0.02}$ & $0.013 \pm 0.004$ & $0.85$ \\
 & $11.0$ & $201$ & $12.71^{+0.05}_{-0.05}$ & $14.80^{+0.14}_{-0.24}$ & $8.53^{+2.44}_{-2.50}$ & $12.76^{+0.02}_{-0.02}$ & $0.009 \pm 0.003$ & $0.94$ \\ \hline
$g$-dropout & $9.4$ & $41,373$ & $11.31^{+0.02}_{-0.03}$ & $13.05^{+0.06}_{-0.05}$ & $7.98^{+1.36}_{-1.32}$ & $11.64^{+0.01}_{-0.01}$ & $0.035 \pm 0.002$ & $6.95$ \\ 
 & $9.6$ & $37,537$ & $11.41^{+0.02}_{-0.02}$ & $13.08^{+0.05}_{-0.04}$ & $7.93^{+1.37}_{-1.32}$ & $11.71^{+0.01}_{-0.01}$ & $0.039 \pm 0.002$ & $6.33$ \\ 
 & $9.8$ & $28,741$ & $11.51^{+0.03}_{-0.03}$ & $13.29^{+0.06}_{-0.05}$ & $7.98^{+1.37}_{-1.35}$ & $11.79^{+0.01}_{-0.01}$ & $0.029 \pm 0.002$ & $4.49$ \\
 & $10.0$ & $23,666$ & $11.63^{+0.02}_{-0.02}$ & $13.36^{+0.05}_{-0.05}$ & $8.00^{+1.33}_{-1.35}$ & $11.88^{+0.01}_{-0.02}$ & $0.031 \pm 0.002$ & $4.40$ \\
 & $10.2$ & $18,333$ & $11.78^{+0.03}_{-0.03}$ & $13.46^{+0.06}_{-0.06}$ & $7.97^{+1.38}_{-1.37}$ & $11.99^{+0.01}_{-0.01}$ & $0.032 \pm 0.002$ & $2.29$ \\
 & $10.4$ & $4,443$ & $11.94^{+0.03}_{-0.03}$ & $13.67^{+0.10}_{-0.08}$ & $8.00^{+1.38}_{-1.39}$ & $12.11^{+0.02}_{-0.02}$ & $0.026 \pm 0.003$ & $2.75$ \\
 & $10.6$ & $1,674$ & $12.22^{+0.04}_{-0.03}$ & $13.86^{+0.13}_{-0.13}$ & $8.01^{+1.35}_{-1.35}$ & $12.33^{+0.03}_{-0.03}$ & $0.028 \pm 0.004$ & $2.04$ \\
 & $10.8$ & $541$ & $12.45^{+0.04}_{-0.04}$ & $14.38^{+0.13}_{-0.22}$ & $8.04^{+1.33}_{-1.38}$ & $12.49^{+0.02}_{-0.02}$ & $0.013 \pm 0.004$ & $0.94$ \\
 & $11.0$ & $145$ & $12.79^{+0.05}_{-0.06}$ & $14.83^{+0.24}_{-0.25}$ & $9.05^{+2.01}_{-2.13}$ & $12.75^{+0.02}_{-0.02}$ & $0.008 \pm 0.004$ & $0.88$ \\ \hline
$r$-dropout & $9.8$ & $6,707$ & $11.45^{+0.07}_{-0.06}$ & $13.07^{+0.12}_{-0.13}$ & $8.55^{+1.74}_{-1.77}$ &  $11.65^{+0.05}_{-0.05}$ & $0.036 \pm 0.005$ & $4.98$ \\
 & $10.0$ & $6,271$ & $11.57^{+0.03}_{-0.03}$ & $13.42^{+0.11}_{-0.11}$ & $8.25^{+2.30}_{-2.18}$ & $11.74^{+0.02}_{-0.02}$ & $0.020 \pm 0.002$ & $4.89$ \\
 & $10.2$ & $4,925$ & $11.72^{+0.03}_{-0.03}$ & $13.62^{+0.10}_{-0.11}$ & $8.28^{+2.29}_{-2.27}$ & $11.85^{+0.02}_{-0.02}$ & $0.017 \pm 0.002$ & $4.98$ \\
 & $10.4$ & $2,806$ & $12.00^{+0.03}_{-0.03}$ & $14.14^{+0.10}_{-0.13}$ & $8.18^{+2.27}_{-2.20}$ & $12.08^{+0.02}_{-0.02}$ & $0.009 \pm 0.002$ & $3.18$ \\
 & $10.6$ & $1,181$ & $12.29^{+0.05}_{-0.04}$ & $14.23^{+0.23}_{-0.30}$ & $8.82^{+2.59}_{-2.61}$ & $12.29^{+0.03}_{-0.04}$ & $0.011 \pm 0.003$ & $2.07$ \\ \hline
\end{tabular}
\label{tab:param}
\end{center}
\end{table*}
\end{turnpage}

\begin{table*}[htb]
\begin{center}
\caption{The Best-Fit HOD Parameters with $1\sigma$ errors of $u$-dropout Galaxies Limited by the Stellar Mass by Varying All of the HOD Free Parameters}
\begin{tabular}{ccccccc} \hline \hline
$\log(M_{{\rm \star, limit}}/\Msun)$ & $\log(\Mmin/h^{-1}\Msun)$ & $\log(M_{1}/h^{-1}\Msun)$ & $\log(M_{0}/h^{-1}\Msun)$ & $\sigmalogM$ & $\alpha$ & $\chi^{2}/{\rm dof}$ \\ \hline
$9.4$ & $11.32^{+0.03}_{-0.04}$ & $13.23^{+0.23}_{-0.30}$ & $8.00^{+1.90}_{-2.06}$ & $0.08111^{+0.01988}_{-0.04580}$ & $0.8907^{+0.2407}_{-0.1186}$ & $4.96$ \\ 
$9.6$ & $11.50^{+0.04}_{-0.03}$ & $13.70^{+0.46}_{-0.33}$ & $8.15^{+2.13}_{-2.11}$ & $0.04049^{+0.01503}_{-0.01866}$ & $0.9791^{+0.2134}_{-0.1930}$ & $6.38$\\
$9.8$ & $11.55^{+0.05}_{-0.03}$ & $13.71^{+0.35}_{-0.34}$ & $8.03^{+2.05}_{-2.10}$ & $0.05911^{+0.02037}_{-0.03020}$ & $0.9260^{+0.2410}_{-0.1534}$ & $5.92$\\
$10.0$ & $11.62^{+0.02}_{-0.04}$ & $13.47^{+0.38}_{-0.29}$ & $8.11^{+2.14}_{-2.13}$ & $0.04425^{+0.01418}_{-0.02438}$ & $0.9737^{+0.2278}_{-0.1883}$ & $5.79$\\
$10.2$ & $11.81^{+0.01}_{-0.02}$ & $13.71^{+0.59}_{-0.20}$ & $8.50^{+2.35}_{-2.40}$ & $0.02057^{+0.00897}_{-0.00650}$ & $1.076^{+0.154}_{-0.250}$ & $5.36$\\
$10.4$ & $11.89^{+0.01}_{-0.01}$ & $13.90^{+0.43}_{-0.32}$ & $8.38^{+2.33}_{-2.30}$ & $0.02673^{+0.01209}_{-0.01104}$ & $0.9768^{+0.2236}_{-0.1942}$ & $5.23$ \\
$10.6$ & $12.14^{+0.01}_{-0.01}$ & $14.13^{+0.28}_{-0.33}$ & $8.64^{+2.35}_{-2.53}$ & $0.01740^{+0.00833}_{-0.00473}$ & $1.075^{+0.151}_{-0.153}$ & $5.12$ \\ \hline
\end{tabular}
\label{tab:param_fix}
\end{center}
\end{table*}

\section{DISCUSSION}
\subsection{HOD Parameters}
\begin{figure}[tbp]
\begin{center}
\epsscale{1.25}
\plotone{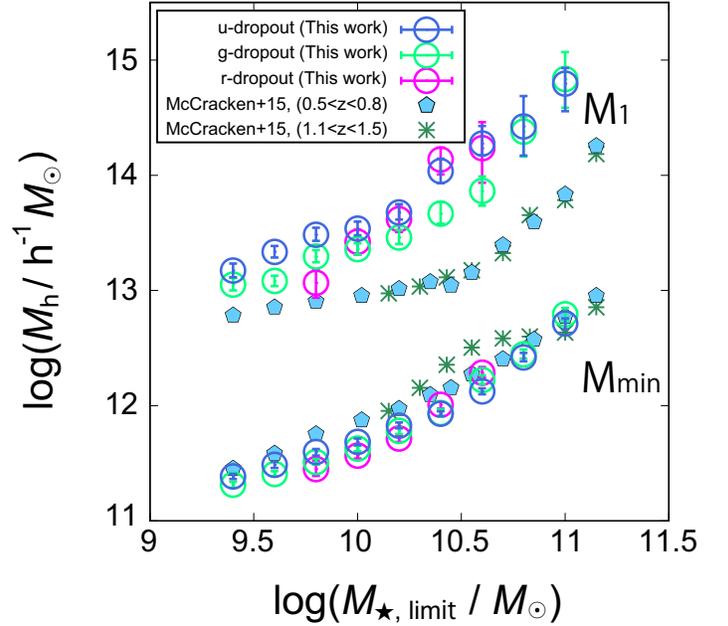}
\caption{Best-fit HOD parameters from the HOD fitting as a function of the stellar-mass limit. The HOD parameters $\Mmin$ and $M_{1}$ represent characteristic halo masses to possess one central galaxy and a satellite galaxy within dark haloes, respectively. The results of \citet{mccracken15} at $0.5 < z < 0.8$ and $1.1 < z < 1.5$ are also shown. }
\label{fig:HODparams}
\end{center}
\end{figure}

One of the advantages of adopting the HOD formalism is that each HOD parameter has an explicit physical meaning; $\Mmin$ and $M_{1} + M_{0}$ are the characteristic halo masses to possess one central and one satellite galaxy within the dark haloes, respectively. 
In Figure~\ref{fig:HODparams}, we show the best-fit values of $\Mmin$ and $M_{1}$ of each subsample as a function of the threshold galaxy stellar mass. 
$\Mmin$ and $M_{1}$ show approximately linear increases with the stellar-mass limit. 
We also plot $\Mmin$ and $M_{1}$ at $0.5 < z < 0.8$ and $1.1 < z < 1.5$ by \citet{mccracken15} based upon the HOD analysis for all galaxies collected by photometric redshifts, although $\Mstar$ of \citet{mccracken15} represents the median stellar mass. 
By comparing the results at low-$z$, the increasing trends of $\Mmin$, i.e., the threshold halo mass to possess a central galaxy, are not significantly different at high-$z$, indicating that the formation efficiencies of the central galaxies are almost the same, at least $z = 5 - 0$. 

However, the behavior of $M_{1}$, which is the threshold halo mass to host a satellite galaxy, is clearly different between $z = 3-5$ and $z = 0-1$. 
First, $M_{1}$ increases linearly with stellar mass from the less massive end to intermediate stellar mass ($\Mstar \sim 10^{10.6} \Msun$), and evolves drastically toward the massive stellar-mass end at low-$z$ Universe. 
This trend is the same as the results of high-$z$; however, the turning stellar mass apparently shifts toward the lower stellar mass compared with low-$z$. 
The rapid increase of $M_{1}$ at high stellar mass indicates that massive satellite galaxies are formed less efficiently than less massive satellite galaxies. 
The lower turnover stellar mass of $M_{1}$ at high-$z$ Universe suggests that dynamical frictions for accreted haloes are not yet sufficient even for less massive haloes. 

In addition, $M_{1}$ are apparently excess for $z = 3-5$ compared with $z = 0-1$, inferring that satellite galaxies only form within massive dark haloes in high-$z$ Universe. 
One possible explanation is that dark haloes at high-$z$ Universe could have higher virial temperature than those in local Universe, because high-$z$ dark haloes are thought have not been completely virialized. 
This could prevent intra-halo gases from cooling and consequently star formation efficiency decreases. 
Another possibility is the effect of feedback from central galaxies. 
Our galaxy samples are confined to star-forming galaxies and galactic winds, if it is more frequent or powerful at high-$z$, from central galaxies may hamper the formation of satellite galaxies. 

\subsection{Mean Halo Masses}
The mean halo mass, $\Mh$, and the satellite fraction, $\fsat$, can be estimated from the best-fit HOD parameters for each subsample. 
The mean halo mass and the satellite fraction are defined as
\begin{equation}
\Mh = \frac{1}{n_{g}}\int dM M n(M) N_{{\rm tot}}(M)
\label{eq:mh}
\end{equation}
and
\begin{equation}
\fsat = 1 - \frac{1}{n_{g}} \int dM n(M) \Ncen(M).
\end{equation}
We note that $n_{g}$ is calculated by equation~(\ref{eq:nghod}). 

\begin{figure}[tbp]
\begin{center}
\epsscale{1.0}
\plotone{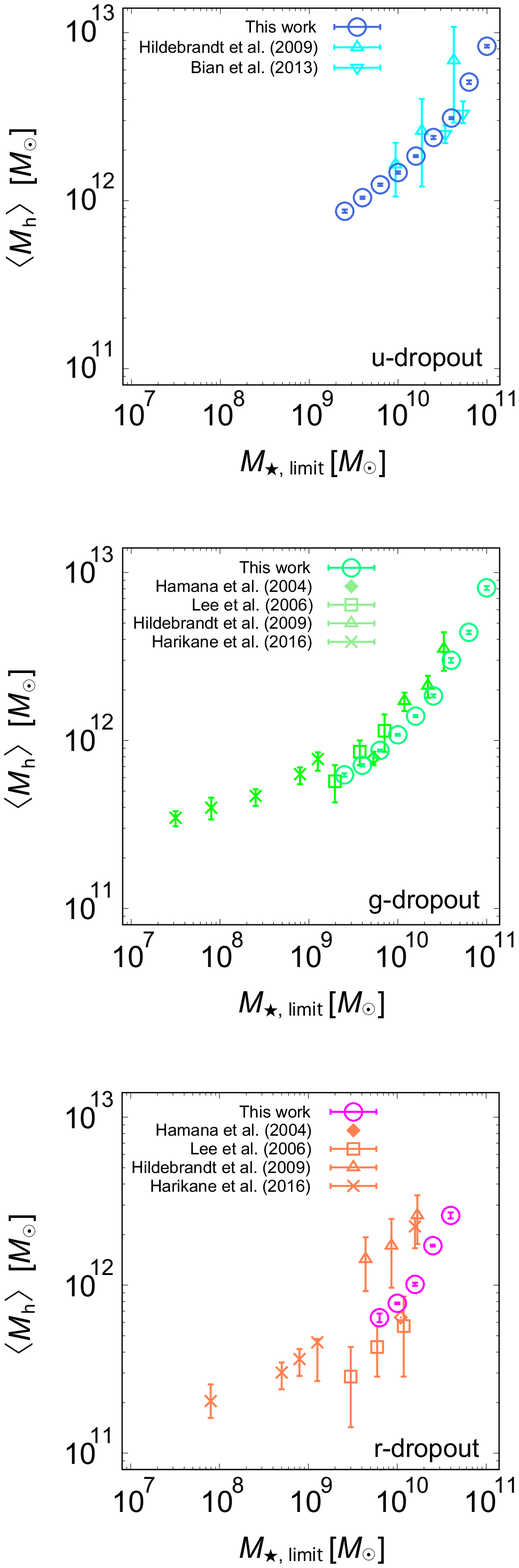}
\caption{Mean halo masses of $u$- (top), $g$- (middle), and $r$-dropout galaxies (bottom) as a function of the stellar-mass limit. For comparison, we also show the mean halo masses of previous HOD studies. Note that the stellar-mass limits of \citet{hamana04}, \citet{lee06}, \citet{hildebrandt09}, and \citet{bian13} are converted from their UV-magnitude limits assuming that the UV slope of $\beta$ follow the observed linear $\beta-M_{{\rm UV}}$ relation at each redshift \citep{bouwens14}. The $\beta-M_{{\rm UV}}$ relation of $z\sim4$ is assumed for converting $u$-dropout galaxies. All masses are in units of $\Msun$, assuming that $h = 0.7$. }
\label{fig:Mhalo}
\end{center}
\end{figure}

The mean halo masses as a function of the stellar-mass threshold of each dropout galaxy are shown in Figure~\ref{fig:Mhalo}. 
The mean halo masses can be derived for wide stellar-mass ranges, especially of $u$-dropout galaxies, by taking advantages of wide and deep images of the CFHTLS. 
The large number of our LBG sample enables us to measure the masses of dark haloes with small error bars, even for very massive haloes ($\Mh \sim 10^{13}\Msun$), although the fixed $\sigmalogM$ and $\alpha$ artificially reduced the errors. 
At a fixed stellar-mass limit, the mean halo masses decrease slightly with increasing redshift. 

We compare the mean halo masses measured by the previous HOD studies for high-$z$ Universe: \citet{hamana04}, \citet{lee06}, \citet{hildebrandt09}, \citet{bian13}, and \citet{harikane16}. 
Our results are almost consistent with these studies within $1\sigma$ confidence level; however, the halo masses of $z \sim 5$ LBGs calculated by \citet{hamana04} and \citet{lee06} were lower than those of \citet{hildebrandt09} and our results. 
It should be noted that other studies adopted the different HOD model, as well as cosmological parameters, from the ones in the present study. 
\citet{hamana04} proposed and used an early halo occupation model that does not distinguish the central galaxy from satellite galaxies, and \citet{lee06}, \citet{hildebrandt09}, and \citet{bian13} also employed the same occupation model. 

It should also be taken into consideration that cosmological parameters are not unified between these studies; all of the referenced works used simple flat ${\rm \Lambda}$CDM cosmology ($\Omega_{{\rm m}} = 0.3$, $\Omega_{{\rm \Lambda}} = 0.7$), whereas we adopt the \textit{WMAP} seven-year cosmologies ($\Omega_{{\rm m}} = 0.27$, $\Omega_{{\rm \Lambda}} = 0.73$). 
The largest impact on the halo mass estimation is the value of $\sigma_{8}$. 
This study and \citet{harikane16} adopt $\sigma_{8} = 0.8$, but other referenced studies used $\sigma_{8} = 0.9$; thus, our results may underestimate the halo mass compared with studies using $\sigma_{8} = 0.9$. 

\subsection{Satellite Fractions}
\begin{figure*}[tbp]
\begin{center}
\epsscale{1.00}
\plotone{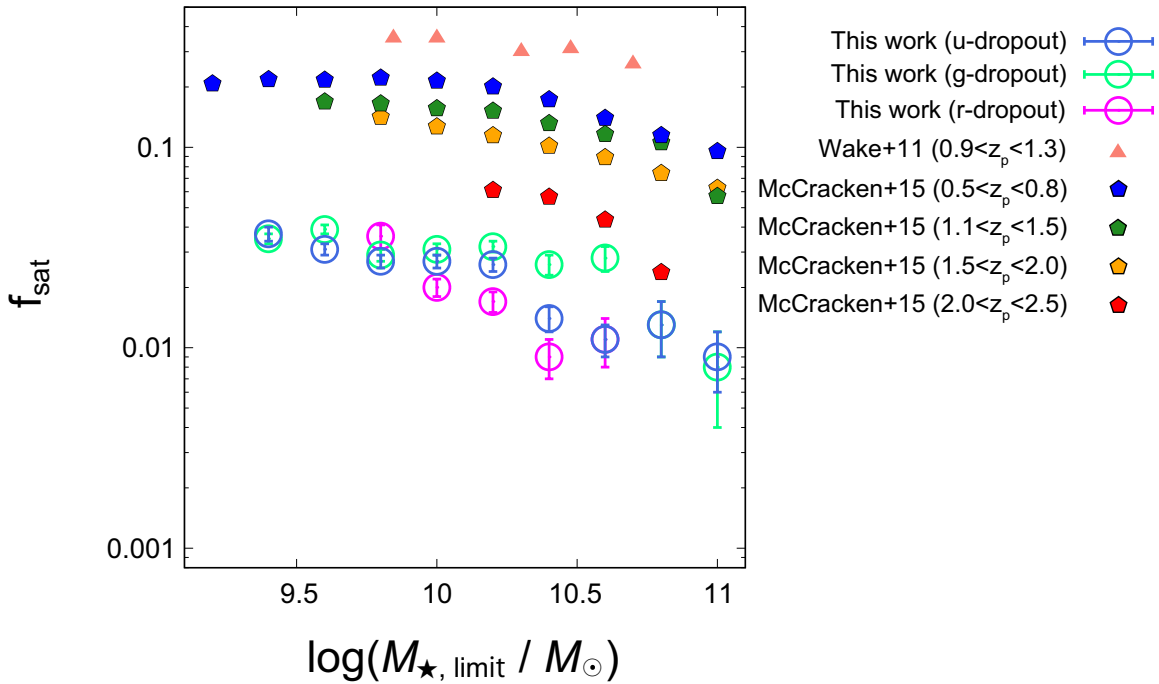}
\caption{Evolution of satellite fractions of $u$-, $g$-, and $r$-dropout galaxies as a function of stellar-mass limit. For comparison, we also plot satellite fractions at $0.5 < z < 2.5$ derived by \citet{wake11} and \citet{mccracken15}. It is noted that our galaxy samples are star-forming galaxies according to their star-formation rates, whereas \citet{wake11} and \citet{mccracken15} selected total galaxy samples using photometric redshifts. }
\label{fig:fsat}
\end{center}
\end{figure*}

In Figure~\ref{fig:fsat}, we show the satellite fractions as a function of the stellar-mass threshold of each dropout galaxy. 
As expected, the satellite fractions decrease gradually with increasing the galaxy stellar mass and then seem to drop sharply by increasing the stellar-mass limit at each redshift. 
The turning stellar mass of the satellite fraction corresponds approximately to the turnover stellar-mass limit of $M_{1}$. 

Our results show that the satellite fractions are quite low, even for less massive galaxies ($\fsat \sim 0.04$ for $M_{{\rm \star, limit}} = 10^{9.4} \Msun$); this is completely different from the results of the $0<z<2$ Universe \citep[e.g.,][]{wake11,zehavi11,coupon12,martinez15,mccracken15}. 
\citet{wake11} investigated the satellite fraction of galaxies at $1 < z < 2$ collected by the SED fitting method and showed that the satellite fraction decreases with increasing the stellar mass; however, the values of $\fsat$ have a large gap compared to our results ($\fsat = 0.21$ for galaxy samples with $\Mstar = 3.0 \times 10^{10} \Msun$ at $\bar{z} = 1.9$). 
\citet{martinez15} carried out HOD analyses for the \textit{Spitzer}/IRAC detected galaxies at $z \sim 1.5$ using the \textit{Spitzer} South Pole Telescope Deep-Field Survey \citep[SSDF;][]{ashby13} and showed that the satellite fraction of their faintest magnitude-limit samples ($[4.5]_{{\rm limit}} = 18.6$ mag with a median stellar mass of $\Mstar = 10^{9.87} \Msun$) was $\fsat = 0.23$. 
They compared the satellite fractions of \citet{zehavi11}, who implemented HOD analyses using numerous local SDSS galaxies, and inferred that the satellite fraction grows slowly from $\fsat \sim 0.2$ at $z \sim 2$ to $\fsat \sim 0.3$ at $z \sim 0$. 
Low-$z$ studies collected all galaxies, whereas our high-$z$ studies can only select star-forming galaxies with high SFRs. 
Thus, we may have missed quiescent high-$z$ galaxies with different baryonic natures. 
Our results indicate that there are few satellite galaxies even for less massive galaxies at high-$z$; the satellite fraction shows little evolution at $z = 3 - 5$ and significant evolution from $z \sim 3$ to $z \sim 2$. 
Accurate satellite fractions in high-$z$ Universe require extensive and complete high-$z$ galaxy survey in the future. 

\subsection{Stellar-to-Halo Mass Ratios}
SHMR is an efficient diagnostic to evaluate the relationship between galaxies and their host haloes, which represents the star-formation efficiency within a dark halo, determined by the integrated star-formation history of the galaxy. 
The SHMR is defined as the fraction between the stellar masses of galaxies and host halo masses, $\Mstar/M_{{\rm h}}$, as a function of a dark halo mass. 

In this section, we compare SHMRs of our LBG samples with previous theoretical and observational studies. 
First, we check the consistency of the SHMRs calculated by different stellar-mass estimation methods, by the MS relation and the SED fitting technique, and then compare our results with literature and discuss the redshift evolution. 
For the comparison, we use the threshold stellar masses of subsamples as $\Mstar$ and $\Mmin$ as $M_{{\rm h}}$ to compute the SHMRs of LBGs, i.e., our SHMRs are computed by $M_{{\rm \star, limit}}/\Mmin$ for each stellar-mass limit subsample. 
The errors of the $\Mmin$ are evaluated by the MCMC simulations during the HOD fitting analyses, whereas the errors of the $M_{{\rm \star, limit}}$ are adopted the root mean squares of the stellar mass of each subsample. 

\subsubsection{SHMRs from main sequence vs. SED fitting}
\begin{figure}[tbp]
\begin{center}
\epsscale{1.35}
\plotone{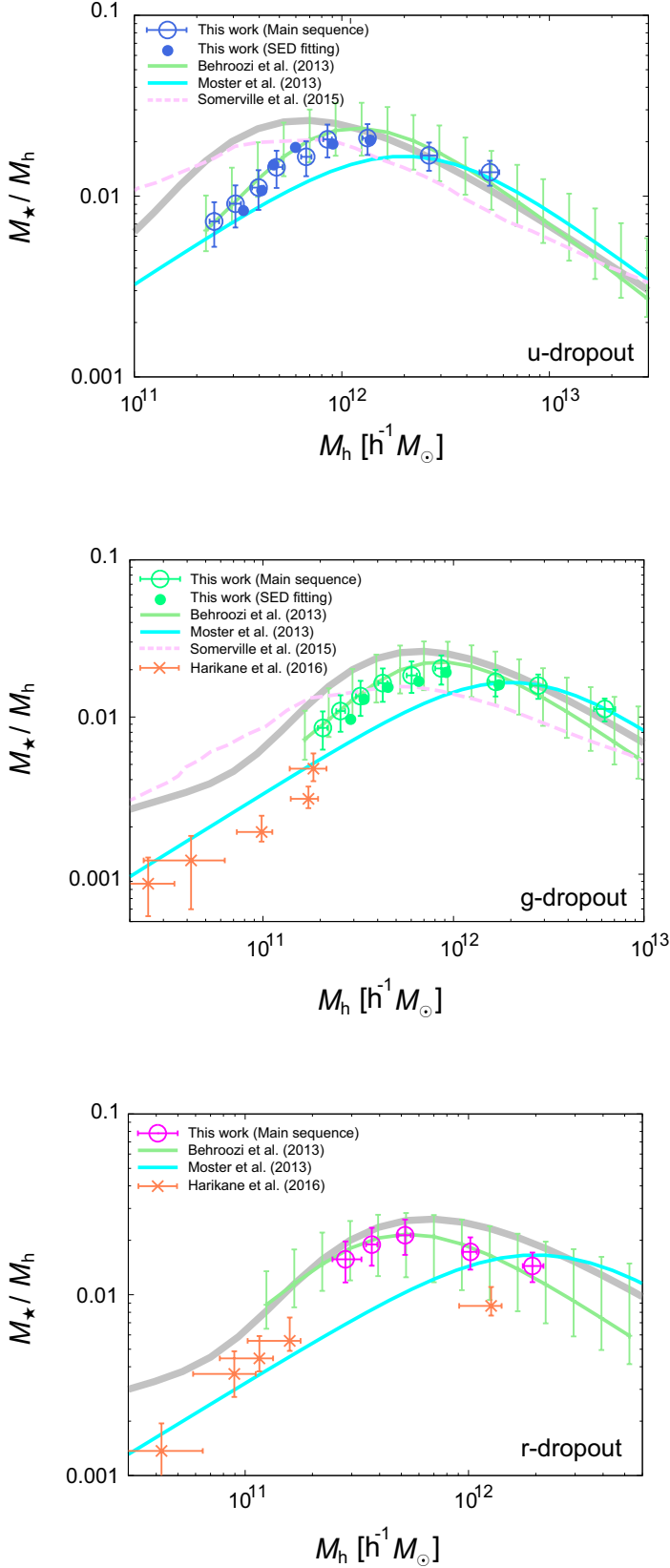}
\caption{Comparison of our SHMRs of $u$- (top), $g$- (middle), and $r$-dropout galaxies (bottom) with literature. The open circles with error bars are our SHMRs computed from the stellar masses by assuming the relation of the main sequence of star-forming galaxies, whereas filled circles (only $u$- and $g$-dropout galaxies) are evaluated via the SED fitting technique. Solid lines are model predictions from the abundance-matching method (green with error bars: \citealt{behroozi13b} at each redshift, gray: \citealt{behroozi13b} at $z\sim0$, and cyan: \citealt{moster13}) and dashed pink line is the results of the semi-analytical model \citep{somerville15}. The observational results of \citet{harikane16} are also shown (orange crosses with error bars). }
\label{fig:SHMR}
\end{center}
\end{figure}

In Figure~\ref{fig:SHMR}, we compare the SHMRs calculated by the MS relation and the SED fitting method of $u$- and $g$-dropout galaxies. 
The SED fitting method, which is difficult to apply reliably to faint samples, i.e., less massive galaxies, has a higher lowest-mass bin than the MS relation, and also shows a relatively large scatter than those from the MS relations due to the small survey volume where NIR data are available. 
The results indicate that the stellar-mass estimated by the SED fitting and the MS relation are not significantly different on the measurement of SHMRs. 
They are consistent within $1\sigma$ error over almost the entire mass range. 
We note that our SED fitting includes the prior of the MS relation (equation \ref{eq:MS}), which is the same relation used to estimate the stellar masses from the MS relation. 
Hereafter, we use the SHMRs derived from the MS relations, which can be applied for wide stellar-mass ranges with high S/N ratios based upon wide-field data. 
We also assume that the consistency is valid for $r$-dropout galaxies, in which the stellar masses from the SED fitting technique are suffered from the relatively large scatter because the Balmer/$4000{\rm \AA}$ break moves beyond the $K$-band for $z \sim 5$ galaxies. 

\subsubsection{Comparison with literature}
In Figure~\ref{fig:SHMR}, our observed SHMRs are compared with the results of numerical simulations by \citet{behroozi13b} and \citet{moster13}, and a semi-analytical model by \citet{somerville15} \citep[GK model;][]{gnedin10}, and observational data from \citet{harikane16}. 
The assignment method of central galaxies with host dark haloes of \citet{behroozi13b} and \citet{moster13} are the abundance-matching method and \citet{harikane16} is the HOD formalism. 
 \citet{behroozi13b} and \citet{moster13} presented their own functional forms of SHMRs as a function of the halo mass and those relations take into account redshift evolutions by being calibrated with the stellar-mass functions at each redshift. 
We plot the predicted SHMRs of each author at $z =3.0$, $4.0$, and $5.0$ in Figure~\ref{fig:SHMR}. 

Our results show an excellent agreement with the SHMRs of \citet{behroozi13b} in all redshift bins, indicating that there is little difference between the HOD technique and the abundance-matching method as a galaxy--halo connection method, at least for the high-redshift Universe. 
In addition to the constraints on the stellar-mass functions, \citet{behroozi13b} also introduced observational constraints, such as cosmic SFR and specific SFRs (sSFRs), which are essential parameters to characterize high-$z$ galaxies. 
For $z = 3$ and $4$, \citet{moster13} well reproduces our observational results only for the massive end but underestimates from the less massive end to intermediate halo mass ($\Mhalo < 10^{12.0} h^{-1}\Msun$ for $M_{{\rm \star, limit}} \sim 10^{10.4} \Msun$) and $\Mpivot$ (see Section~5.4.3.) shifts toward higher mass. 
This discrepancy may be caused by that \citet{moster13} could not constrain the SHMRs for less massive galaxies by stellar-mass functions, whereas \citet{behroozi13b} could reproduce it up to $\Mstar \sim 10^{9.4} \Msun$, which is comparable to our stellar-mass limit. 
The functional form of \citet{moster13} cannot describe our observational results at $z = 5$ because their adopted stellar-mass functions \citep{pg08,santini12} do not reach $z = 5$. 
The $\Mpivot$ of \citet{somerville15}, on the other hand, show lower halo masses compared to observation. 
This may be due to the overproduction of less massive galaxies in their model. 
We note that our SHMRs are inconsistent with the theoretical prediction of \citet{yang12}, who calculated the SHMR based upon the conditional luminosity function \citep[CLF;][]{yang03,vdbosch03b}. 

\citet{harikane16} observed the SHMRs for very less massive LBGs at $z = 4$ and $5$. 
Their results are not conflict to our results, although their estimates show slightly lower SHMRs than ours. 
This may partly be due to differences in the employed cosmological parameters, or because they introduced a fixed duty cycle of the star formation in their halo occupation model; by introducing the fixed duty cycle parameter, the expected number of LBG (or total amount of stellar masses) for fixed halo masses change, possible affecting the values of the SHMR and $\Mmin$. 

\subsubsection{Evolution of the SHMRs and the pivot halo masses}
A halo mass with the maximum SHMR, a pivot halo mass $(\Mpivot)$, can be interpreted as the most efficient dark halo mass to convert baryonic contents into stars within dark haloes. 
\citet{leauthaud12} firstly observed a decreasing trend of $\Mpivot$ from $z = 1$ to $z = 0.2$ by joint analyses of galaxy--galaxy lensing, galaxy spatially clustering, and the stellar mass function in the COSMOS field. 
They also showed that maximum SHMRs are almost consistent, albeit the decrease in $\Mpivot$, indicating an increase in star-forming activities in low-mass haloes with decreasing redshift, known as galaxy downsizing. 

\citet{mccracken15} evaluated $\Mpivot$ up to $z \sim 2$ via the abundance-matching technique in addition to the HOD analysis, and compared their results with those in literature \citep{coupon12,leauthaud12,hudson15,martinez15}. 
They concluded that $\Mpivot$ does not significantly evolve from $z = 0$ to $z \sim 2$ as a consequence of the little evolution of halo mass functions and stellar mass functions up to $z \sim 2$. 

Due to the availability of deep and wide field imaging data of the CFHTLS, our sample can, for the first time, trace the evolution of the pivot halo mass at $3 < z < 5$. 
We fit our SHMRs using the parameterized form proposed by \citet{behroozi10} as follows: 
\begin{equation}
\log(\Mstar(M)) = \log(\epsilon M_{1}) + f\biggl( \log\biggl(\frac{M}{M_{1}}\biggr) \biggr) - f(0), 
\label{eq:SHMR_behroozi}
\end{equation}
where $f(x)$ satisfies the following relationship: 
\begin{equation}
f(x) = -\log(10^{\alpha x} + 1) + \delta \frac{(\log(1+ \exp(x)))^{\gamma}}{1 + \exp(10^{-1})}. 
\label{eq:SHMR_f}
\end{equation}
In this parameterized function, $M_{1}$ and $\epsilon$ represent the pivot halo mass and the SHMR at the pivot halo mass, respectively. 
We investigated the best-fit parameters of $M_{1}$ and $\epsilon$ for each dropout galaxy with fixing $\alpha$, $\delta$, and $\gamma$ for their best-fit values at each redshift presented in \citet{behroozi13b}. 
The $\Mpivot$ and SHMRs at $\Mpivot$ of each dropout galaxy are summarized in Table~\ref{tab:Mpivot}. 

\begin{figure}[tbp]
\begin{center}
\epsscale{1.20}
\plotone{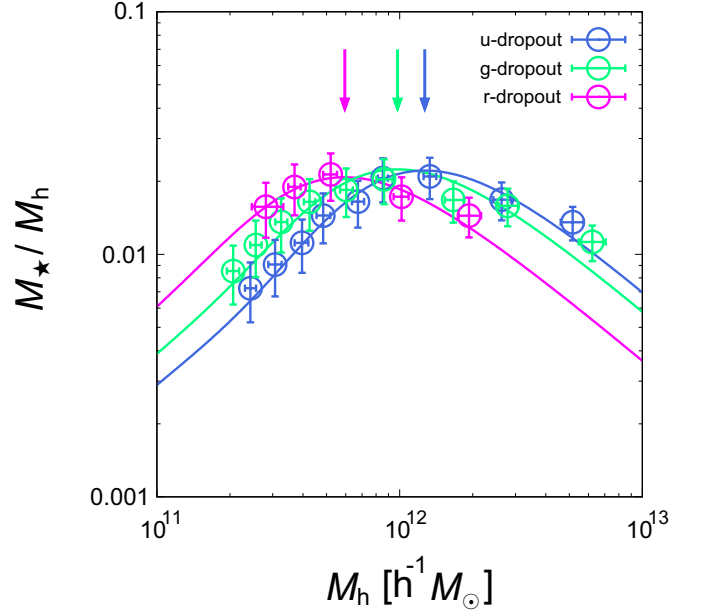}
\caption{SHMRs of $u$- (blue), $g$- (green), and $r$-dropout galaxies (red). Solid lines are the best-fit functions of SHMRs fitted by the parameterized function proposed by \citet{behroozi10}. Arrows above the SHMRs indicate the pivot halo masses of each dropout galaxy. }
\label{fig:comp_SHMR}
\end{center}
\end{figure}

Figure~\ref{fig:comp_SHMR} shows the SHMRs and best-fit functions of each redshift. 
At a fixed halo mass, our SHMRs show the negative evolution for galaxies within less massive dark haloes ($\Mhalo \lesssim 10^{12.0}h^{-1}\Msun$) and the positive evolution for massive haloes with decreasing redshift. 
The arrows above the SHMRs indicate the best-fit $\Mpivot$ of each dropout sample. 
First of all, $\Mpivot$ at $z=3-5$ are measured to be $\sim 10^{12}h^{-1}\Msun$, which is comparable to local Universe. 
The $\Mpivot$ slightly increases but the SHMRs at $\Mpivot$ are not significantly changed at $z = 3 - 5$, which is consistent with the observational results of LBGs at $z=4-7$ \citep{stefanon16}. 
The efficiency of the star formation changes very little from $z=5$ to $z=0$, which is consistent with time-independent star-formation efficiency model \citep{behroozi13b}. 
\citet{fakhouri10} reported the mean mass accretion rates as $\langle \dot{M}_{{\rm halo}}\rangle_{{\rm mean}} \propto (\Mhalo / 10^{12}\Msun) \times (1+z)^{2.5}$ at $z \gtrsim 1$ based upon the results of the cosmological $N$-body simulations \citep{springel05,boylan09}. 
LBGs show active star-forming activities that is comparable to the above rapid growth of dark halo masses in high-redshift Universe. 
Combined with the low satellite fractions, these highly star-forming activities may not be governed by galaxy mergers. 

\begin{figure}[tbp]
\begin{center}
\epsscale{1.20}
\plotone{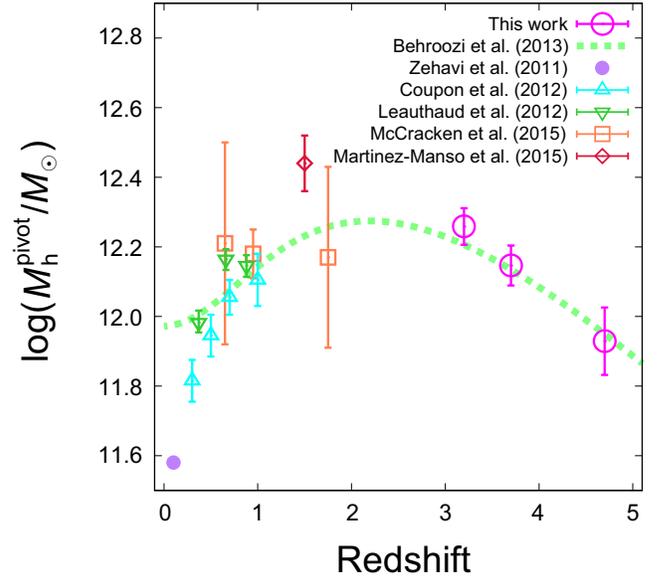}
\caption{Pivot halo masses calculated by HOD analyses as a function of redshift. The dashed green line indicates the model prediction of the pivot halo mass evolution calculated by the equation (\ref{eq:SHMR_behroozi}) and (\ref{eq:SHMR_f}) \citep{behroozi10}. Pivot halo masses are given in units of $\Msun$ on logarithmic scale by assuming $h =0.7$. }
\label{fig:Mpivot}
\end{center}
\end{figure}

$\Mpivot$ shows the small positive evolution with cosmic time, which is opposite to previous lower-$z$ studies at $z = 0 - 1$ \citep[e.g.,][]{coupon12,leauthaud12}. 
In Figure~\ref{fig:Mpivot}, we compare the $\Mpivot$ with previous HOD studies and a model prediction of the pivot halo mass evolution by \citet{behroozi10} calculated by equation (\ref{eq:SHMR_behroozi}) and (\ref{eq:SHMR_f}). 
We show observationally, for the first time, the increasing trend of $\Mpivot$ with cosmic time at $z > 3$; this $\Mpivot$ evolution is well described by the model prediction using the abundance-matching technique. 
$\Mpivot$ evolves from $\Mpivot=10^{11.77}h^{-1}\Msun$ at $z \sim 5$ to $\Mpivot=10^{12.10}h^{-1}\Msun$ at $z \sim 3$, which can be described as $\log{(\Mpivot/\Msun)} \propto -0.22 \times (1+z)$ for this redshift range, and little evolution of SHMRs at $\Mpivot$ indicates that stellar-mass growth is comparable to halo mass growth in high-$z$ Universe. 
This idea is in agreement with the observational results of the rapid stellar-mass evolution from $z > 3$, when intense halo assembly is expected, toward the era of the cosmic noon \citep[e.g.,][]{hopkins06,suzuki15}. 
In contrast, at $z < 2$, previous HOD studies show an increasing trend of $\Mpivot$ with increasing redshift. 
This trend is evidence for galaxy downsizing, i.e., galaxy star-formation is taking place within less massive haloes with cosmic time from $z \sim 2$. 

More interestingly, $\Mpivot$, which is the halo mass with the most efficient star-formation activity, is almost unchanged around $\log(\Mpivot/\Msun) = 12.1 \pm 0.2$ by comparing with the local values as well as the model prediction \citep{behroozi13b} over cosmic time at $0 < z < 5$. 
This suggests that the peak in the star-formation efficiency and its normalization are almost time independent, while environment of $10^{12} \Msun$ haloes has significantly evolves since $z=5$. 
This has been theoretically predicted by considering of the shock heating, the supernova feedback, the gas cooling, and the gas accretion \citep[e.g.,][]{rees77,white78,dekel06}. 
Our study shows an observational evidence that the pivot halo mass around $10^{12}\Msun$ represents the characteristic dark halo mass for galaxy formation. 

\begin{table}[htb]
\begin{center}
\caption{Best-fit Values of the Pivot Halo Mass and the SHMRs at Pivot Halo Mass}
\begin{tabular}{lcc} \hline \hline
 & $\log(\Mpivot/h^{-1} \Msun)$ & $\Mstar / \Mpivot (\times 10^{-2})$ \\ \hline
$u$-dropout & $12.10 \pm 0.053$ & $2.222 \pm 0.189$ \\ 
$g$-dropout & $11.99 \pm 0.057$ & $2.248 \pm 0.190$ \\ 
$r$-dropout & $11.77 \pm 0.097$ & $2.091 \pm 0.228$ \\ \hline
\end{tabular}
\label{tab:Mpivot}
\end{center}
\end{table}

\section{CONCLUSIONS}
We present the results of the clustering and HOD analyses of Lyman break galaxies at $z = 3$, $4$, and $5$. 
Thanks to the deep- and wide-field images of the CFHTLS Deep Survey, we collected a large number of LBGs, enabling us to derive sufficiently accurate ACFs to apply the HOD analysis to see and discuss the relationship between LBGs and their host dark haloes even in high-redshift Universe with high accuracy. 
Stellar masses of sample galaxies, resulting from either the MS relation or the SED fitting technique, helped us to link high-$z$ dropout galaxies to their host dark haloes. 

The major conclusions of this work are summarized below. 
\begin{enumerate}
\item The clustering amplitude of each dropout galaxy depends on the galaxy stellar mass. 
We reveal that more massive LBGs in stellar mass show stronger clustering amplitude, which is consistent with the results of the previous LBG clustering study \citep{hildebrandt09} and the low-$z$ studies \citep[e.g.,][]{mccracken15}. 

\item Characteristic halo masses possessing a central and a satellite galaxies ($\Mmin$ and $M_{1}$) show approximately linear growth at $z = 3 - 5$. 
$\Mmin$ follows the same increasing trend with stellar masses at low-$z$, whereas $M_{1}$ at $z = 3 - 5$ shows higher values than $z = 0.5 - 1.5$ over the entire stellar-mass range, suggesting that satellite galaxies are formed inefficiently within dark haloes at high-$z$. 
This implies that the high virial temperature and/or the strong feedback effect from the central galaxy in high-$z$ dark haloes may prevent effective satellite galaxy formation. 
In addition, the $M_{1}$ increases more sharply than at low-$z$. 
This discrepancy of $M_{1}$ dependence on the stellar mass may be due to the short time-scale to accrete galaxies for being satellite galaxies. 

\item The mean halo masses calculated by the best-fit HOD model are $\Mh = 10^{11.7}h^{-1}\Msun \sim 10^{12.8}h^{-1}\Msun$, which are consistent with previous HOD studies. 
The satellite fractions of dropout galaxies are less than $0.04$ for our stellar-mass limit, indicating that satellite galaxies rarely form in high-redshift dark haloes. 
By comparing the satellite fractions at $z = 0-2$, it is expected the drastic evolution of the number of satellite galaxies from $z \sim 3$ to $z \sim 2$, albeit the difference of selection criterion among galaxy populations (e.g., the galaxy stellar mass and the SFR). 

\item The SHMRs are computed using two independent stellar mass estimate methods: the SED fitting technique and by assuming the main sequence of star-forming galaxies. 
We confirm that these two estimations agree with each other. 

\item The observed SHMRs at $z = 3 - 5$ are in good agreement with the model prediction of \citet{behroozi13b}. 
We derive observationally, for the first time, the pivot halo mass at $3 < z < 5$, which shows a slight increasing trend with cosmic time at $z > 3$. 
In contrast, the values of SHMRs for pivot halo masses show little evolution. 
This suggests that the mass growth rates of stellar components and dark haloes are comparable at $3 < z < 5$. 

\item $\Mpivot$ is found to be almost unchanged around $\log(\Mpivot/\Msun) = 12.1 \pm 0.2$ over cosmic time at $0 < z < 5$, suggesting that the peak in the star-formation efficiency and its normalization are almost time independent, while environment of $10^{12} \Msun$ haloes has significantly evolves since $z=5$. 
This has been theoretically predicted by considering of the shock heating, the supernova feedback, the gas cooling, and the gas accretion. 
The pivot halo mass at $\Mhalo \sim 10^{12}\Msun$ can be, so to speak, the characteristic dark halo mass for galaxy formation. 
\end{enumerate}

\acknowledgments
We thank to Frank van den Bosch for giving us useful comments. 
NK acknowledges supports from the JSPS grant 15H03645. 
KI acknowledges supports from the Grant-in-Aid for JSPS fellow for young researchers (PD). 
This study is based on observations obtained with MegaPrime/MegaCam and WIRCam, a joint project of CFHT and CEA/DAPNIA, at the Canada--France--Hawaii Telescope (CFHT), which is operated by the National Research Council (NRC) of Canada, the Institut National des Sciences de l'Univers of the Centre National de la Recherche Scientifique (CNRS) of France, and the University of Hawaii. 
This work is based in part on data products produced at TERAPIX and the Canadian Astronomy Data Center as part of the Canada--France--Hawaii Telescope Legacy Survey and WIRCam Deep Survey, a collaborative project of NRC and CNRS.
Data analyses of this work were in part carried out on common use data analysis computer system at the Astronomy Data Center, ADC, of the National Astronomical Observatory of Japan. 
This work was partially supported by Overseas Travel Fund for Students (2016) of the
Department of Astronomical Science, SOKENDAI (the Graduate University for Advanced
Studies). 

{\it Facilities:} \facility{CFHT (MegaCam and WIRCam)}.

\end{document}